# Urban spatial-temporal activity structures: a New Approach to Inferring the Intra-urban Functional Regions via Social Media Check-In Data

Ye Zhi[1,2,3,4], Yu Liu[1], Shaowen Wang[2,3,4], Min Deng[5], Jing Gao[2,3,4], Haifeng Li[2,3,4,5]*

[1] Institute of Remote Sensing and Geographical Information Systems, Peking University, Beijing, China

[2] Department of Geography and Geographic Information Science, University of Illinois at Urbana-Champaign, Urbana, IL, USA

[3] CyberGIS Center for Advanced Digital and Spatial Studies, University of Illinois at Urbana-Champaign, Urbana, IL, USA

[4] National Center for Supercomputing Applications (NCSA), University of Illinois at Urbana-Champaign, Urbana, IL, USA

[5] Department of Surveying and Geo-informatics, Central South University, Changsha, China

**Abstract:** Most existing literature focuses on the exterior temporal rhythm of human movement to infer the functional regions in a city, but they neglects the underlying interdependence between the functional regions and human activities which uncovers more detailed characteristics of regions. In this research, we proposed a novel model based on the low rank approximation (LRA) to detect the functional regions using the data from about 15 million check-in records during a yearlong period in Shanghai, China. We find a series of latent structures, called urban spatial-temporal activity structure (USTAS). While interpreting these structures, a series of outstanding underlying associations between the spatial and temporal activity patterns can be found. Moreover, we can not only reproduce the observed data with a lower dimensional representative but also simultaneously project both the spatial and temporal activity patterns in the same coordinate system. By utilizing the K-means clustering algorithm, five significant types of clusters which are directly annotated with a corresponding combination of temporal activities can be obtained. This provides a clear picture of how the groups of regions are associated with different activities at different time of day. Besides the commercial and transportation dominant area, we also detect two kinds of residential areas, the developed residential areas and the developing residential areas. We further verify the spatial distribution of these clusters in the view of urban form analysis. The results shows a high consistency with the government planning from the same periods, indicating our model is applicable for inferring the functional regions via social media check-in data, and can benefit a wide range of fields, such as urban planning, public services and location-based recommender systems and other purposes.

# 1. Introduction

Understanding the distribution of different functional regions (e.g. residential, business and transportation areas, etc.) in a city is a lasting theme in urban studies [1-3]. For many years, exploring the functional regions in the intra-urban area mainly relies on socio-demographics data and aggregate areas with high economic interaction [5]. However, the process of updating such data is labor intensive and time consuming [4], so the results are often stagnant and cannot reflect the dynamic nature of local urban areas.

Recently, as the rapid growth of Information and Communications Technology (ICT), the Location Aware Devices (LAD), including smart phone and Global Positioning System (GPS) enabled cars, are widely used, large quantities of human movement data are collected . A branch of research have tried to utilize the spatial-temporal patterns of human movements to infer the functional regions on a large scale using mobile phone Erlang data [6,7], taxicab data [8,9], smart card data [10,11] and Wi-Fi data [12]. Differing from the movement-based data, some other studies adopted the activity-based survey data to explore the spatial-temporal activity patterns and then illustrates the functional regions of a city [13,14]. Unfortunately, these two kinds of data have their own limitations. The activity-based survey data also requires long hours of observation and causing tremendous time and financial cost [4,7]. Moreover, the outcomes usually do not scale and only uncover a partial depiction of characteristics of functional regions [3]. For the movement-based data, as lack of the travel demand information, they cannot depict detailed characteristics of regions (e.g. the temporal variation of travel demands), thus each cluster of regions' type are inferred by the empirical analysis leading to the non-home/work activities are hard to be distinguished [15]. To overcome these limitations, Yuan et al. simultaneously considers both point of interests (POIs) data of a region and human mobility data to identify functional regions [2]. They infer the user's travel demand by linking the movement with a POI data set, however, this is not always conformed to the real case. For instance, one user goes to a shopping mall next to a campus. If the shopping mall doesn't exist in the POI data set, his/her movement will be linked to education purpose rather than shopping.

Fortunately, with the proliferation of social media, such as Facebook, Twitter, Foursquare and Flickr, millions of registered members are recording their surroundings and sharing their movement routes with friends via check-in [16]. Being different from cell phone and car trajectories data which are derived from GPS trackers, check-in data not only contain the location but also record user's travel demand. Although there exits false check-in (i.e. one user is not actually at airport, however, he/she pretended to create his/her check-in location at airport), Cheng et al. have announced a series of rules to filter out the false check-in records [17]. Wu et al. also proposed five criteria to eliminate the fake check-ins and trips [18]. As a consequence, check-in data has the superior ability to depict the intra-urban functional regions than the previous three kinds of data. Justin et al. have discovered the sub-urban areas (called Livehoods) from check-in data [3]. Thiago et al. also utilized check-in data to measure eight cities dynamics in a large scale [19]. But they does not systematically examine the interdependence between the functional regions and human activity via check-in data.

In addition, since human behaviors in urban involve a wide range of demands with time dimension, one region's characteristics could be delineated as a synthesis of temporal rhythm of multiply activities in a high dimensional matrix [6,20]. However, such matrix contains redundant activity structures because some activities display a strong correlation with others (e.g. dining in the evening usually has a strong association with the entertainment in the evening). As a result, the original high dimensional matrix supposed to be reduced as a lower-dimensional representation. Such lower-dimensional representation represent the inherent structure of the original high dimensional matrix and has a unique signature to reflect the link between functional region and human activity. Most existing research have applied the principal component analysis (PCA) to explore such lower-dimensional representation and then analyzed the eigen structures to interpret the interdependence between the functional regions and the human activity [21-23]. Nathan et al. suggested that the human movement pattern could be represented as a repeating structure, termed eigenbehaviors [24]. Moreover, Jonathan and Francesco et al. have connected eigenbehaviors with functional regions study and used the term, eigenplace, to identify the recurring patterns of urban dynamic [6,12]. However, limited by the covariance matrix, PCA must analyze the spatial or the temporal characteristics as two separate sets of features rather than projecting them simultaneously into the same subspace directly showing the connections between the functional regions and human activities.

In this paper, we proposal a novel model via a low rank approximation method (LRA) to infer the intra-urban functional regions according to the data from about 15 million check-in records during a yearlong period in Shanghai, China. The advances of this study lie in three folds.

First, we found a series of latent structures, called urban spatial-temporal activity structure (USTAS), which could represent the outstanding underlying associations between the functional regions and human activities. While interpreting these subspaces, we can not only reproduce the observed data with a lower dimensional representative but also simultaneously project both the regions and activities in the same coordinate system.

Second, we utilize the USTAS to generate the clusters of regions. There is not any predefined functional region classification to be superimposed on the observations. Thus, we provide a clear picture of how groups of regions are associated with different travel demands at different time of day in the city.

Finally, we further verify the spatial distribution of the clusters of regions in the view of urban form analysis. The results shows a high consistency with the latest government planning, indicating our model is applicable for inferring the functional regions via social media check-in data and will benefit a wide range of fields, such as urban planning, public services and location-based recommender systems and other purposes.

The rest of this paper is organized as follows. The next section introduces the check-in data and study area. The third section describes the method how to find USTAS via low-rank approximation. The result is presented in the fourth section, and conclusions and discussions are given in the fifth section.

## 2. Dataset and study area

In this research, we investigated about 15 million social media check-in records during a yearlong period from September 2011 to September 2012 in Shanghai, the largest Chinese city by population [25]. These records have been applied for modeling the intra-urban human mobility [18], and are also the partial dataset for uncovering the inter-urban trip and spatial interaction [21]. Considering both computational efficiency and the heterogeneous distribution of check-in data points, we chose the central part of city ($50 \times 35 km^2$) as the study area and divided it into 1km×1km square lattices (Figure 1). Moreover, we grouped the travel demands into six categories: home (H), transportation (Tr), work (W), dining (D), entertainment (E) and other (O) since some check-in demand-tags signify the similar demand.

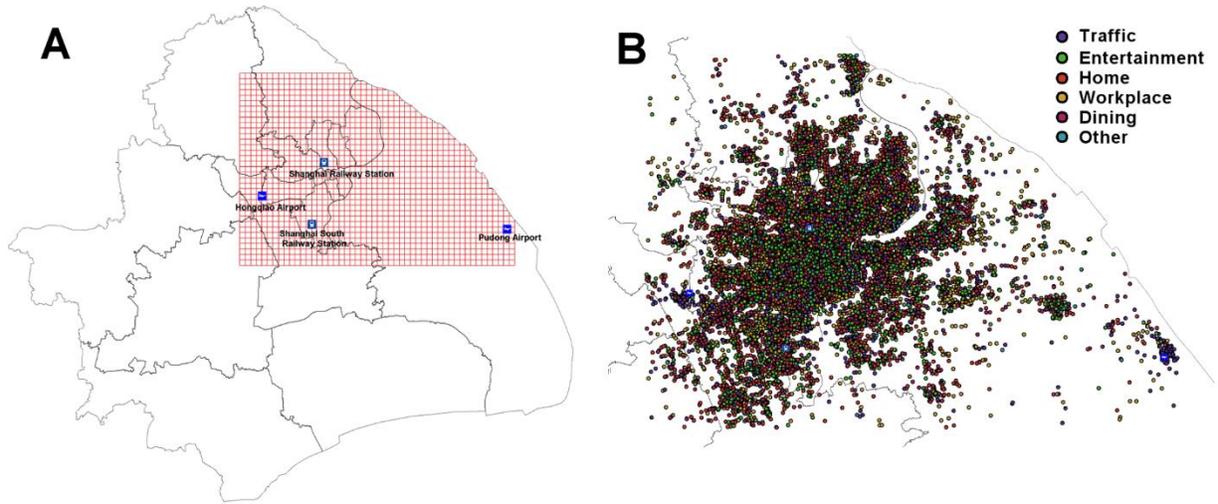

**Figure 1. Spatial distribution of check-ins and the study area.** (a) The red lattices represent the study area, and also involves two airports, Pudong airport and Hongqiao airport, and two railway stations, Shanghai railway station and Shanghai South railway station. (b) Spatial distribution of check-ins by activities. One check-in record is geo-referenced as one point according to its location. Different colors of the points denote different activities.

## 3. Modeling Framework

In this paper, our proposed model mainly includes four steps as the Figure 2 shows: first, we construct a region and temporally-dependent travel demands matrix and then we use a LRA method to find the best estimation of the original matrix; Third, we extract the USTAS from the low rank approximation matrix; Last, we adopt K-means clustering algorithm to aggregate regions into several significant types according to the Top thirty USTAS. We will explain each step as following in details.

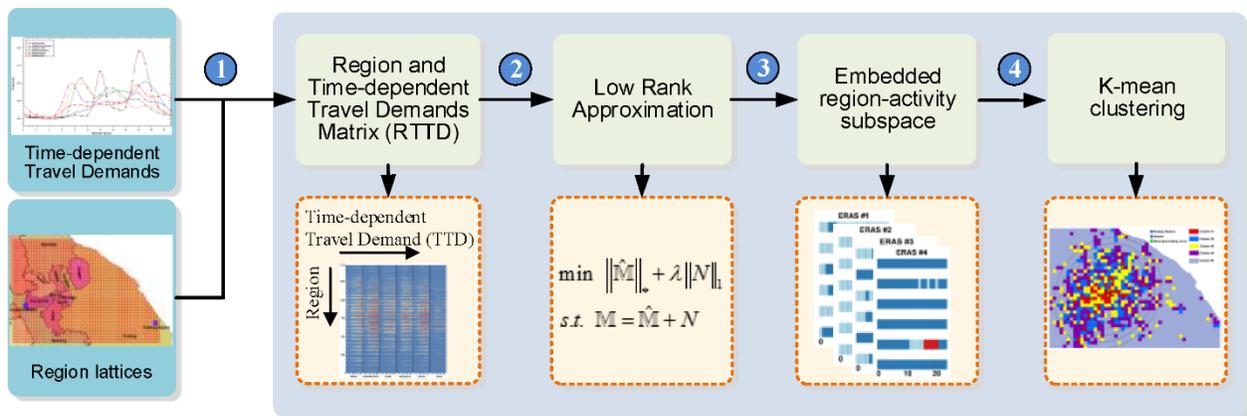

**Figure 2. The modeling framework**

## 3.1 Constructing the Matrix of Region and Temporally-dependent Travel Demands

This paper focuses on exploring the underlying relationship between functional regions and human activity. Thus, we have three basic variables, i.e., $D=\{d_1, d_2, \cdots, d_j\}$ to denote the domain of travel demands, where $j$ means total types of demands, for instance, in this research there are six categories: H, Tr, W, D, E, O; $T=\{t_1, t_2, \cdots, t_k\}$ denote the collection of time intervals set, where $k$ means total temporal intervals in one day, for example, in this research, $k=24$; and $G=\{g_1, g_2, \cdots, g_m\}$ denote the domain of the urban area set, where $m$ means number of square lattices in the previous section. Each square lattices are called as sub-region and could be marked with a certain number ranging from left to right and from bottom to top.

The human activity demands are temporally-relevant, for instance, dining at noon is not the same demand as dining in the evening from the aspect of semantic, and will also cause a different impact on the travel behavior. Therefore, we use Cartesian product of travel demands and time intervals set to denote such the temporally-dependent travel demands (TTD) which can be written as:

$$TTD = D \times T = \{td_1, td_2, \cdots, td_n\}, n = j \times k \qquad (1)$$

The sub-region of $G$ over TTD form a region and TTD relation matrix is denoted by R-TTD or $\mathbb{M}$ for short.

$$\mathbb{M} = (u_{g,td})_{m \times n} \qquad (2)$$

Where $u_{g,td}$ is the intensity of the TTD in the sub-region, meaning the frequency of occurrence for travel demand $d$ at time $t$ conditioned on the sub-region $g$.

## 3.2 Exploring the Low-dimensional representation via LRA

In order to explore the lower-dimensional representation and then analyzed the eigen structures to interpret the interdependence between the functional regions and the human activity, we thus adopt a Low-Rank Approximation (LRA) method. LRA method has been widely applied in the fields of information retrieval [26], face recognition [27] and salient object detection [28], because this approach can explore the latent structure between two associated factors in the high dimensional matrix. The LRA is given as follows:

Any matrix can be decomposition as following two parts [29]:

$$\mathbb{M} = \hat{\mathbb{M}} + N \tag{3}$$

where $\hat{\mathbb{M}}$ is a low rank approximation matrix of $\mathbb{M}$, and $N$ is a perturbation matrix which expresses noise. The best rank-k estimate of $\hat{\mathbb{M}}$ can be denoted as following:

$$\begin{aligned} &\text{minimize} \quad \left\| \mathbb{M} - \hat{\mathbb{M}} \right\|_F \\ &\text{s.t.} \quad rank\left(\hat{\mathbb{M}}\right) \leq k \end{aligned} \tag{4}$$

Then, the matrix $\hat{\mathbb{M}}$ can be decomposed to three matrices according to the Singular Value Decomposition (SVD) method:

$$\hat{\mathbb{M}} = \hat{U}\hat{S}\hat{V}^T \tag{5}$$

where $\hat{U}$ is a $m \times m$ unitary matrix (i.e. $\hat{U}^T\hat{U} = \hat{U}\hat{U}^T = I_{m \times m}$), $\hat{V}$ is a $n \times n$ unitary matrix (i.e., $\hat{V}^T\hat{V} = \hat{V}\hat{V}^T = I_{n \times n}$,), and $\hat{S}$ is a diagonal matrix constraining the singular values $\sigma_i$ of $\hat{\mathbb{M}}$. Since, the $\hat{\mathbb{M}}$ can be also written as the sum of rank-1 matrices:

$$\mathbb{M} \cong \hat{\mathbb{M}} = \sum_{i=1}^{r} \hat{\sigma}_i \hat{u}_i \hat{v}_i^T \tag{6}$$

where $\sigma_i$ is larger than $\sigma_{i+1}, i=1,\cdots,r$, and $r$ equals to the number of nonzero singular values (it also equals to the rank of $\hat{\mathbb{M}}$).

In order to construct $\hat{\mathbb{M}}$, the value of $r$ should be determined. It is suggested to use the Frobenius norm to evaluate the similarity between the $\hat{\mathbb{M}}$ and the original matrix $\mathbb{M}$ as following:

$$E = \frac{\left\| \hat{\mathbb{M}} - \mathbb{M} \right\|_F}{\left\| \mathbb{M} \right\|_F} \times 100\% \tag{7}$$

Where $\|M\|_F = \sqrt{\sum_{i=1}^{m}\sum_{j=1}^{n}\left|a_{ij}^2\right|} = \sqrt{trace\left(A^T A\right)} = \sqrt{\sum_{i=1}^{\min\{m,n\}} \sigma_i^2}$.

## 3.3 Uncovering the urban spatial-temporal activity structure

Based on the Equation 6 and Equation 7, the LRA based on the SVD could be viewed as a transform projecting a high dimension matrix to a series of low-dimensional subspaces. These subspaces can been regarded as multiple intrinsic feature spaces which embedded in the original high dimension data space. Hence, we coin this kind of subspaces as the urban spatial-temporal activity structure (USTAS). One USTAS is viewed as the combination of the columns in the matrix $\hat{U}$ and the columns in the matrix $\hat{V}$ correspondingly (e.g. The combination of $\hat{U}$ 's first column and $\hat{V}$ 's first column could be viewed as the first USTAS). Considering each column in matrix $\hat{U}$ is orthogonal to the other, so one column in $\hat{U}$ denotes one unique feature among regions, called USTAS for the spatial characteristics (USTAS-SC). Similarly, each column in matrix $\hat{V}$ is also orthogonal to the other and one column in $\hat{V}$ could be viewed as one unique feature among TTD, called USTAS for TTD characteristics (USTAS -TTDC). As a result, one USTAS has an ability to simultaneously express the relationship between RC and TTDC.

### 3.4 Identifying the Territory of Functional regions

This step aggregates similar formal regions in terms of USTAS by performing the K-means clustering algorithm. In particular, each row of $\hat{U}$ indicates the characteristics distribution of original sub-region in USTAS and each row of $\hat{V}$ denotes the characteristics distribution of original TTD in USTAS. As a consequence, this result could provide us an opportunity to simultaneously project the sub-regions and TTD into the same subspace for identifying the territory of functional regions, meaning that the $i, j$ cell of the $\hat{\mathbb{M}}$ can be obtained by taking the dot produce of the $i$ and $j$ row of matrix $\hat{V}\hat{S}$ which is denoted as following:

$$\hat{\mathbb{M}} = \hat{U}\hat{S}^{1/2} \left( \hat{V}\hat{S}^{1/2} \right)^T \quad (8)$$

Where $\hat{U}\hat{S}^{1/2}$ is the new coordinate for the original sub-regions and $\hat{V}\hat{S}^{1/2}$ is the new coordinate for the original TTD.

Then we use the K-means clustering method to aggregate similar sub-regions and TTD. Therefore, one cluster indicates one functional region which directly has two kinds of characteristics, the spatial distribution and the function characteristics (the set of TTD in the cluster). K-means clustering method is one of the most fundamental clustering method and has derived a bunch of methods [30], such as K-medoid, K-SVD [31] (Duda, Hart & Stork (2001)). However, two problems exist in the K-means method. One is how to define the distance, and the other one is how to choose the optimal number of clusters.

For the first one, since the basic idea of this work is based on trait representation by vectors in the feature space, hence, we use the cosine distance to measure the dissimilarity of those relationships which is a common method to measure the similarity between two vectors. For the latter one, we should estimate the best cluster numbers for the K-means method, i.e. find the optimal clusters number to be fit for the inherent partition of the data. The most common approaches used to validate the clustering results are: (1) external criteria, whose basic idea is to use previous knowledge about data as external reference; (2) internal criteria whose basic idea is based on only quantities and features intrinsic to the data alone without introduce any prior knowledge about data; and (3) relative criteria, whose basic idea is to choose the best clustering scheme according to a pre-specified criterion without any statistical tests[32].

## 4. Results and analysis

In this work, we set the number of demand categories is $|D|=6$, and the number of time intervals is $|T|=24$ since one hour intervals are adopted as the temporal unit for analysis. The study area has been divided into 1km×1km squares, and the total number of grids is $|G|=1474$ after filtering out water areas.

### 4.1 The visualization of the Region-TTD matrix

In order to give a concrete description of the data set, we visualized the region-travel demands matrix $\mathbb{M}$ as Figure 7 shows. The value of cell, for example, (E15, 760) equals to 493, means the frequency of occurrence for travel demand E in the 15th time interval (i.e. from 14:00 to 15:00) within region 760 is 493. From the view of spatial distribution, some grids show high significance in multiple TTD (e.g. nearby grid 750). On the contrary, some grids are characterized with single TTD (e.g. nearby grid 1250 or grid 250) or have no obvious features (e.g. nearby grid 0 or grid 1474). In the term of temporal characteristics, some TTD display a strong correlation with others (both the frequencies for Home and Entertainment are very high after 7pm while Traffic and workplace are relatively high in the day time). Such characteristics indicate that the matrix $\mathbb{M}$ contains redundant structures and a lower-dimensional representation is suggested to be existing.

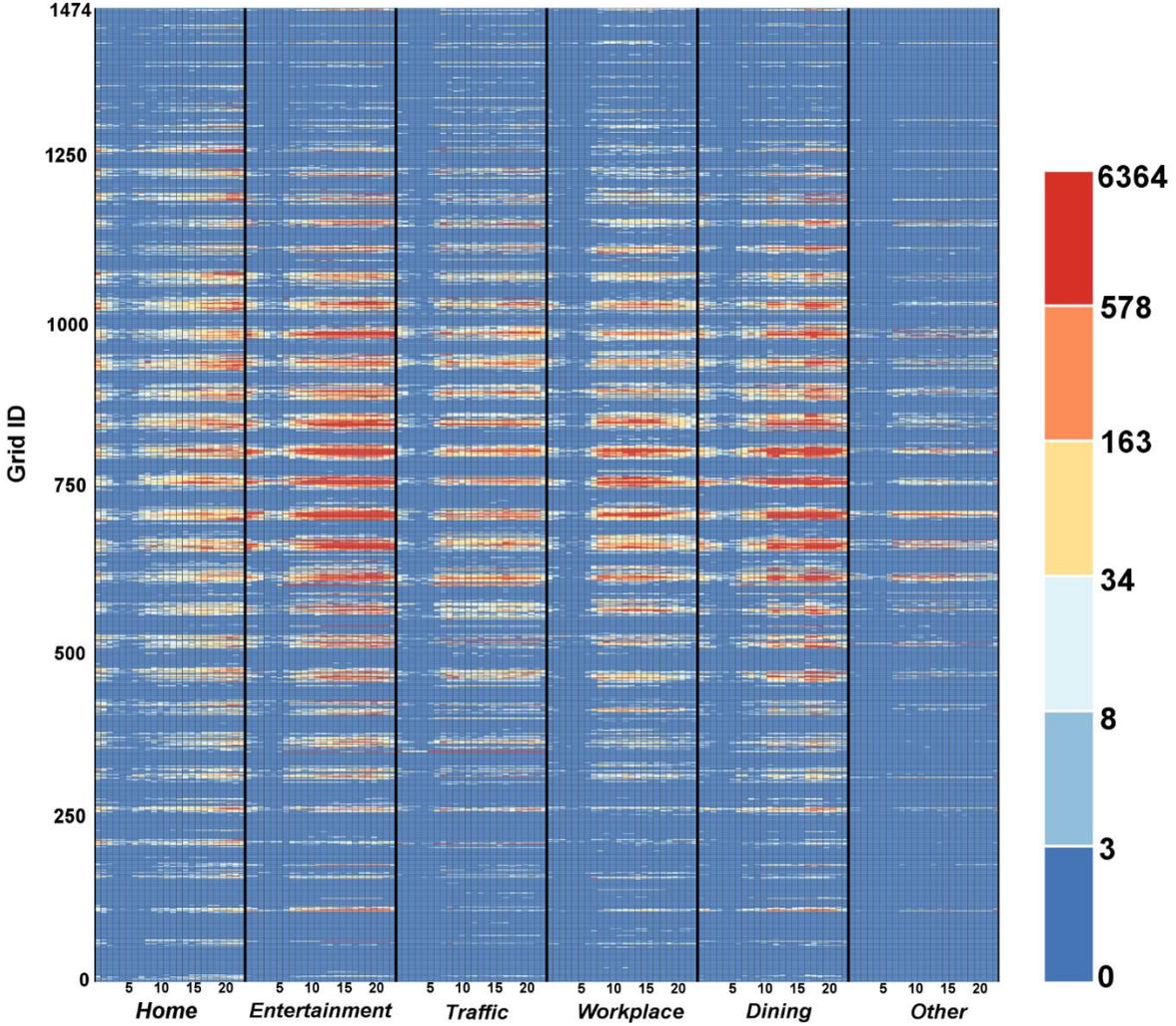

**Figure 3. The Region-TTD matrix .** The vertical axis is the six demands in 24 hours and the vertical axis is the ID sequence of sample grids. This figure illustrates the compounding functions of each grid. Some grids mainly expose one kind of demand, the frequencies of other demands are relatively low while some grids keep high frequency among all of the demands.

## 4.2 The lower-dimensional representation of Region-TTD matrix

In order to explore a lower-dimensional representation $\hat{\mathbb{M}}$ comparing to the original matrix $\mathbb{M}$, based on the Equation 7, it is suggested to determine an optimal $r$. We consider three aspects to choose the value of $r$: the distribution of singular value, the reconstruction accuracy comparing to $\mathbb{M}$, and the reconstruction accuracy comparing to the original temporal variation of travel demands.

**The distribution of singular value**

We set the normalized singular value (ratio to the maximum) as the vertical axe, and index of singular with order from high value to low one (Figure 4). As shown in Equation 7, the $\hat{\mathbb{M}}$ can be written as the sum of $r$ rank-1 matrix. From the view of significance, that most of the feature could be represented by the first few principal structures. The less singular value we need, the more notable feature we have. The Figure 4 shows that the first 30 singular value contribute about ninety percent of energy in $\mathbb{M}$. This phenomenon also tell us that the original matrix $\mathbb{M}$ includes a large number of redundant information which cover up the most valuable and important information about the relationship between the functional region and human activity in the urban area.

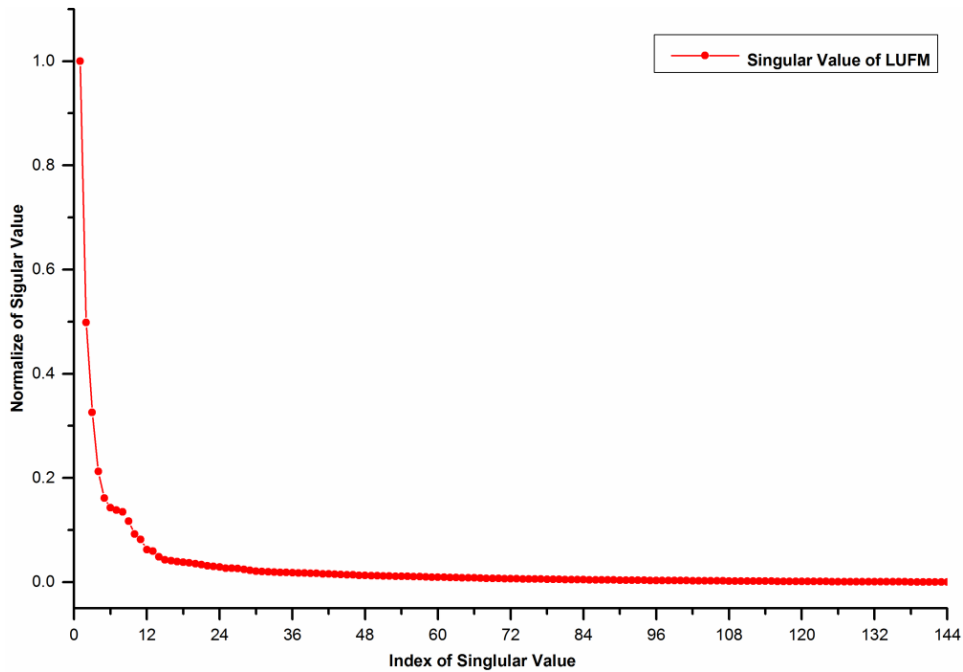

**Figure 4. The distribution of singular value.**

**The reconstruction accuracy comparing to $\mathbb{M}$**

Based on the Equation 7, we can calculate the dissimilarity between the reconstruct matrix $\hat{\mathbb{M}}$ and original matrix $\mathbb{M}$ using different values of $r$ as Figure 5 shows. The horizontal axe means the number of $r$ and the vertical axe means the reconstruction error. It shows that when the value of $r$ equals to 30, the reconstruction error only reaches to 0.05.

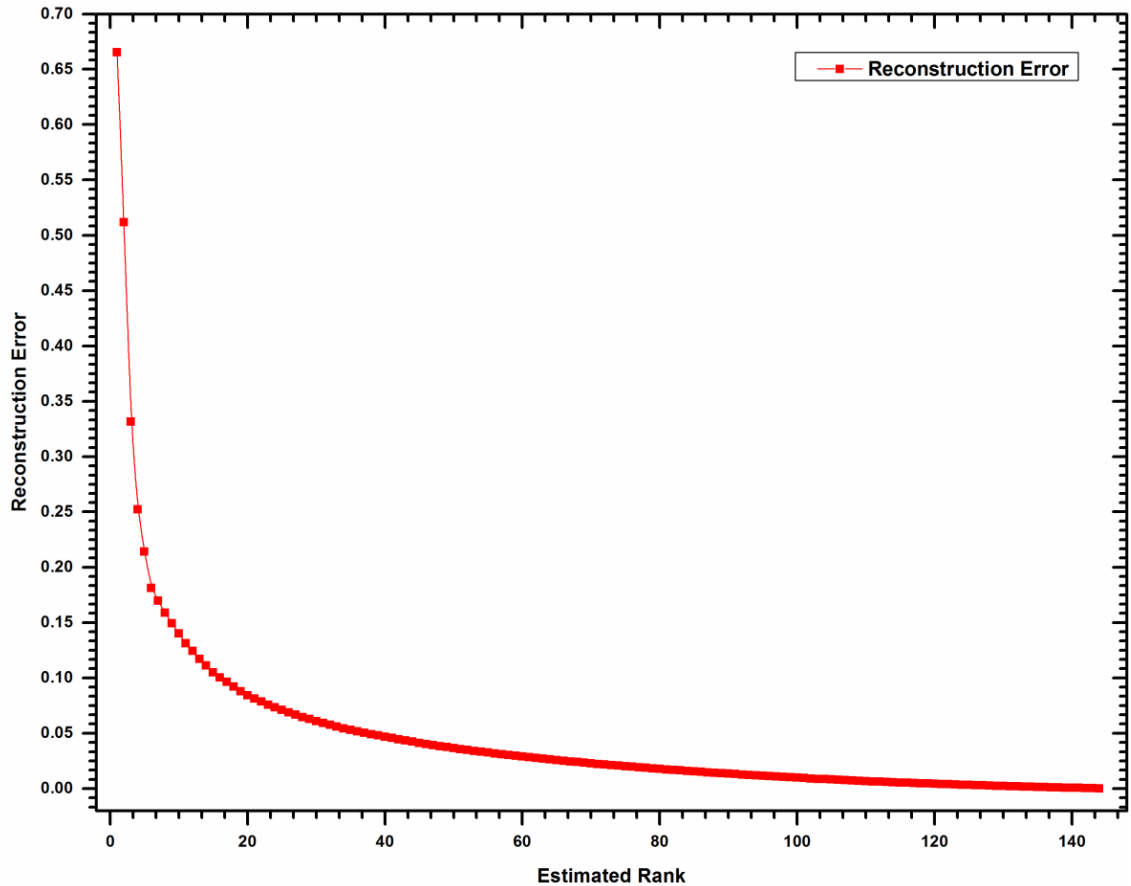

**Figure 5. The reconstruction accuracy comparing to the original matrix**

**The reconstruction accuracy comparing to the original temporal variation of travel demands**

These two above methods indicate that it is appropriate to set the value of $r$ to 30. Then we want to show the difference between the reconstruct TTD and original TTD when r equals to 30. As Figure 6 shows, the horizontal axe means the time intervals in one day, and vertical axe means the frequency of travel demands. We use different color to distinguish the travel demands and different line to discriminate the reconstructed temporal variation of demands from the original ones (dash line for the reconstruction and solid line for the origin). The Figure 6 shows that how well the original distribution of TTD can be approximated when setting the value of $r$ to 30.

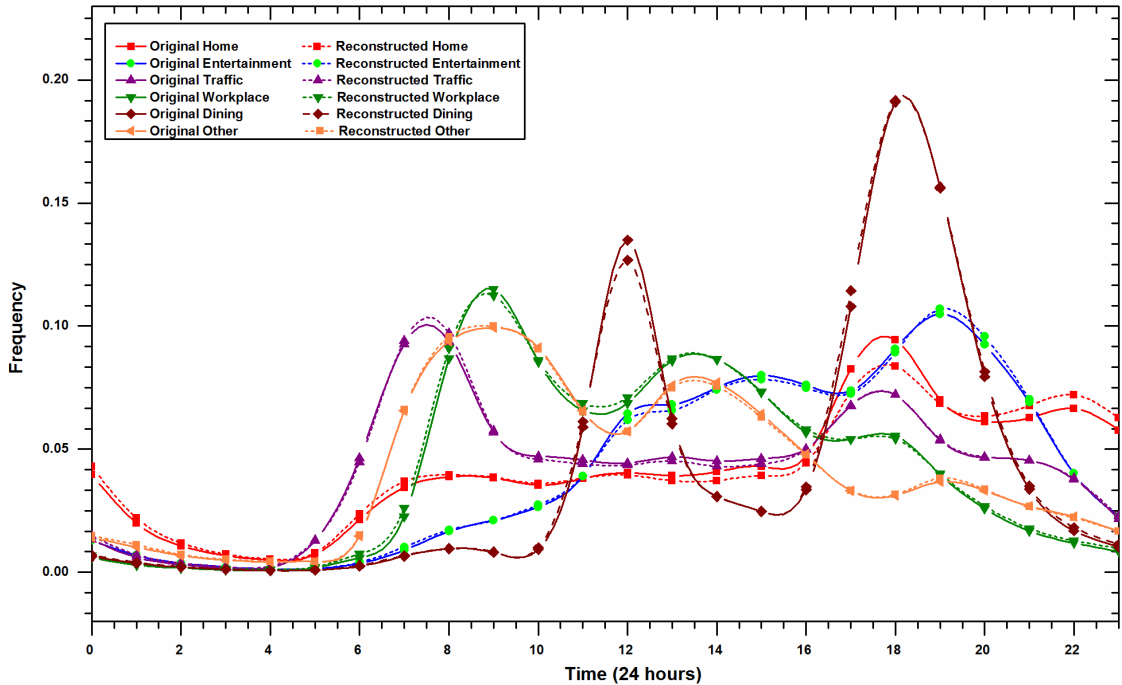

**Figure 6. The reconstruction accuracy comparing to the original temporal variation of travel demands.**

## 4.3 Interpreting the embedded region-activity subspace

In this work, it is suggested that the value of $r$ equals to thirty meaning that there are thirty USTAS to represent the link between the functional region and human activity. Each USTAS consists of two kinds of structures, the spatial structure (one of the columns in $\hat{U}$) and the TTD structure (one of the columns of $\hat{V}$). To have a better understanding of USTAS, we take the top six USTAS as an example. The Figure 8 illustrates the TTD characteristics of the top six USTAS. Correspondingly, the Figure 9 shows the spatial characteristics of the top six USTAS (the green points are the ten municipal trading circles which are obtained from A Collection of Policies for Development in Shanghai Commercial Sector 2010[1]). The first USTAS-TTDC (represented with the first column of six figures in Figure 8) denotes the pattern showing remarkable compounding characteristics of two activities, including the activity for entertainment from noon to 22 pm and the activity for dining at noon and in the evening. Meanwhile, we can get spatial distribution insights from the Figure 9a that the first USTAS-SC mainly accumulated in the most of municipal trading circles, especially in the Nanjing East Road and Huaihai Middle Road which are famous for the pedestrianized tourist street. Comparing to the first USTAS, the second one has lower correlation to the entertainment while shows higher correlation to three kinds of temporal activities involving the activity for traffic in the morning and evening rush hours, the work in the daytime and the activity for dining from noon to 22pm. Although the spatial distribution of the second USTAS is also near the trading circles, it mainly represents workspaces of the trading circles rather than the entertainment parts. The latter three USTAS just show the feature with single activity. For example, the fifth USTAS significantly correlated to the transportation stations in Shanghai, including two airports (Hongqiao airport and Pudong airport) and two railway stations (Shanghai Railway Station and Shanghai South Railway Station). The sixth USTAS mainly relates to the activity for Home corresponding to the resident areas in Shanghai as Figure 9f shows.

---

[1] http://images.mofcom.gov.cn/shanghai/accessory/201006/1277190604257.pdf

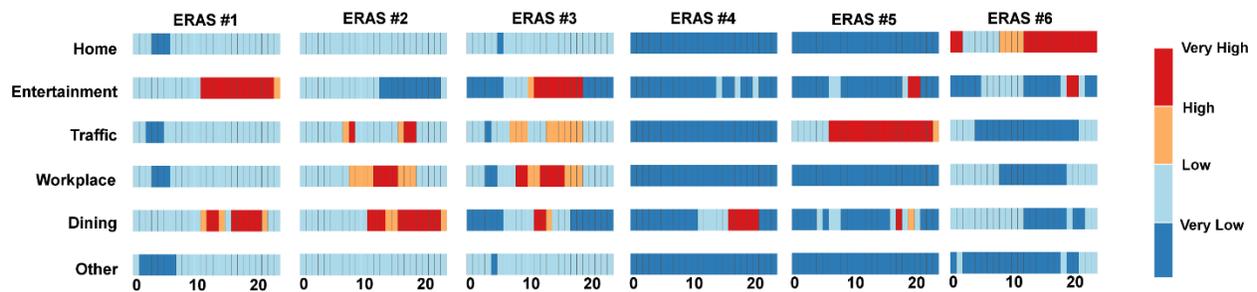

**Figure 7. The activity characteristics of the top six USTAS.** The vertical axis represents the travel demand and horizontal axis implies the time from 0 to 23 during a day. The first three USTAS present significant compounding temporal activity patterns while the latter three USTAS mainly shows some specific single activity patterns.

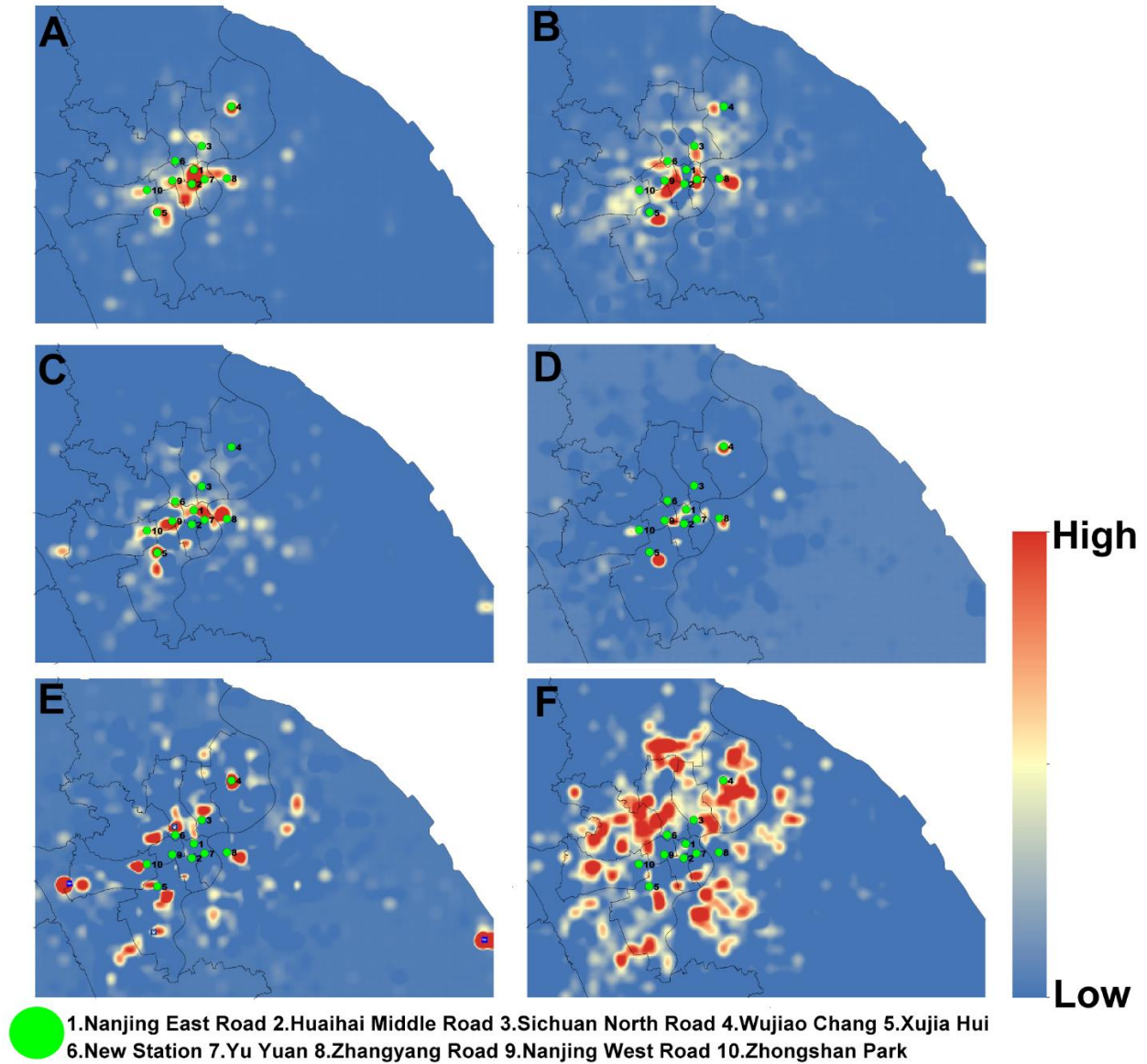

**Figure 8. The spatial characteristics of the top six (USTAS).** The green circles indicate the ten municipal trading circles. The first four USTAS mainly accumulated in the central part while the latter two USTAS expose the discrete space out of the central part involving the transportations stations and the resident areas, respectively.

## 4.4 Inferring the functional regions

In order to demonstrate the importance of USTAS, we applied the USTAS into inferring the functional regions. Referencing the top 30 USTAS, we simultaneously project both the spatial and TTD structures in the same coordinate system. Based on the Equation 7, we use the K-means clustering method to aggregate similar sub-regions and TTD. Without any prior knowledge and pre-specified clustering numbers，we used three typical internal validation criteria, Dunn's index (Dunn 1973), Davies-Bouldin index (DBI) [33] and Silhouette index (Rousseeuw 1987) to determine the optimal number of clusters as the Figure 9 shows. The results show that the optimal number of clustering groups is five.

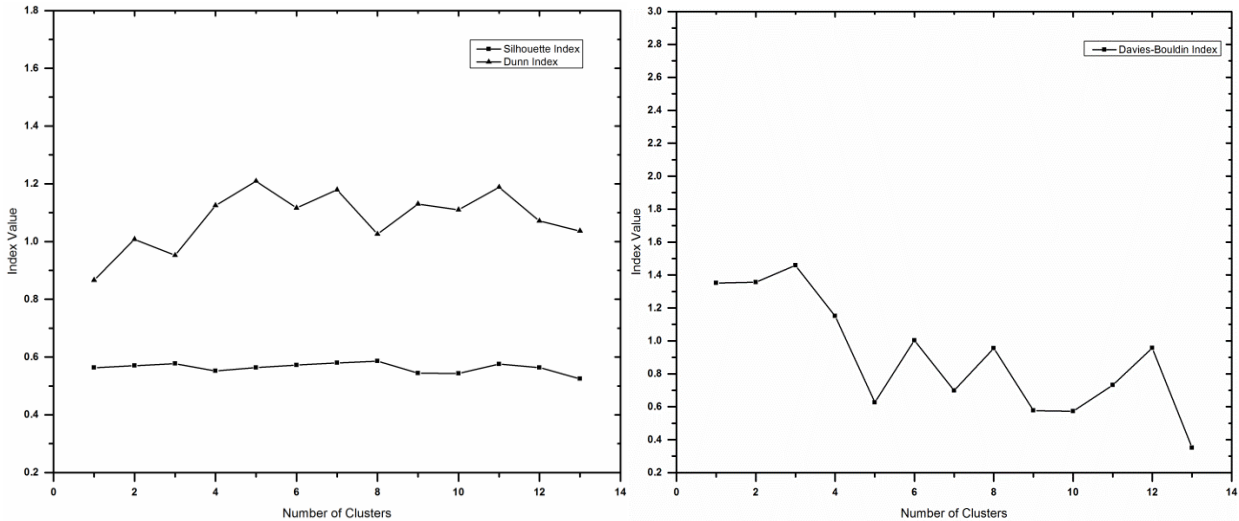

**Figure 9. The Cluster validity indices of the K-means Method**

As a consequence, we found five types of clusters which present significant characteristics in spatial and temporally-dependent travel demands as the Figure10 and Figure11 (a) show. The cluster #1 presents a high degree of the Entertainment and Dining at noon and in the evening. From the spatial aspect, the cluster #1 involves all the municipal trading circles. As a result, we suggest the cluster #1 is commercial-dominant (CD). Differing from the CD, the cluster #2 is highly correlated to the demand for Transportation and also contains some important transportation stations including airports and railway stations in Shanghai, so the cluster #2 is treated as the transportation-dominant (TD). Both the cluster #3 and cluster # 4 have a strong association with the demand for Home, however, they are dissimilar to the features for other demands and the spatial distribution. All the values of the other demands in cluster #3 are positive, meaning that the activities for the other demands in cluster #3 are also active. By the contrast, the values of the other demands in cluster #4 are negative indicating that the other demands is not associated with the cluster #4. In the view of spatial distribution, the cluster #3 is closer to the CD than the cluster #4. Therefore, we suggested the cluster #3 is the developed residential-dominant (developed RD) and the cluster #4 is the developing residential-dominate (developing RD). Since the cluster #5 shows a low degree of all the demands, we suggested the cluster #5 is other-dominate (OD).

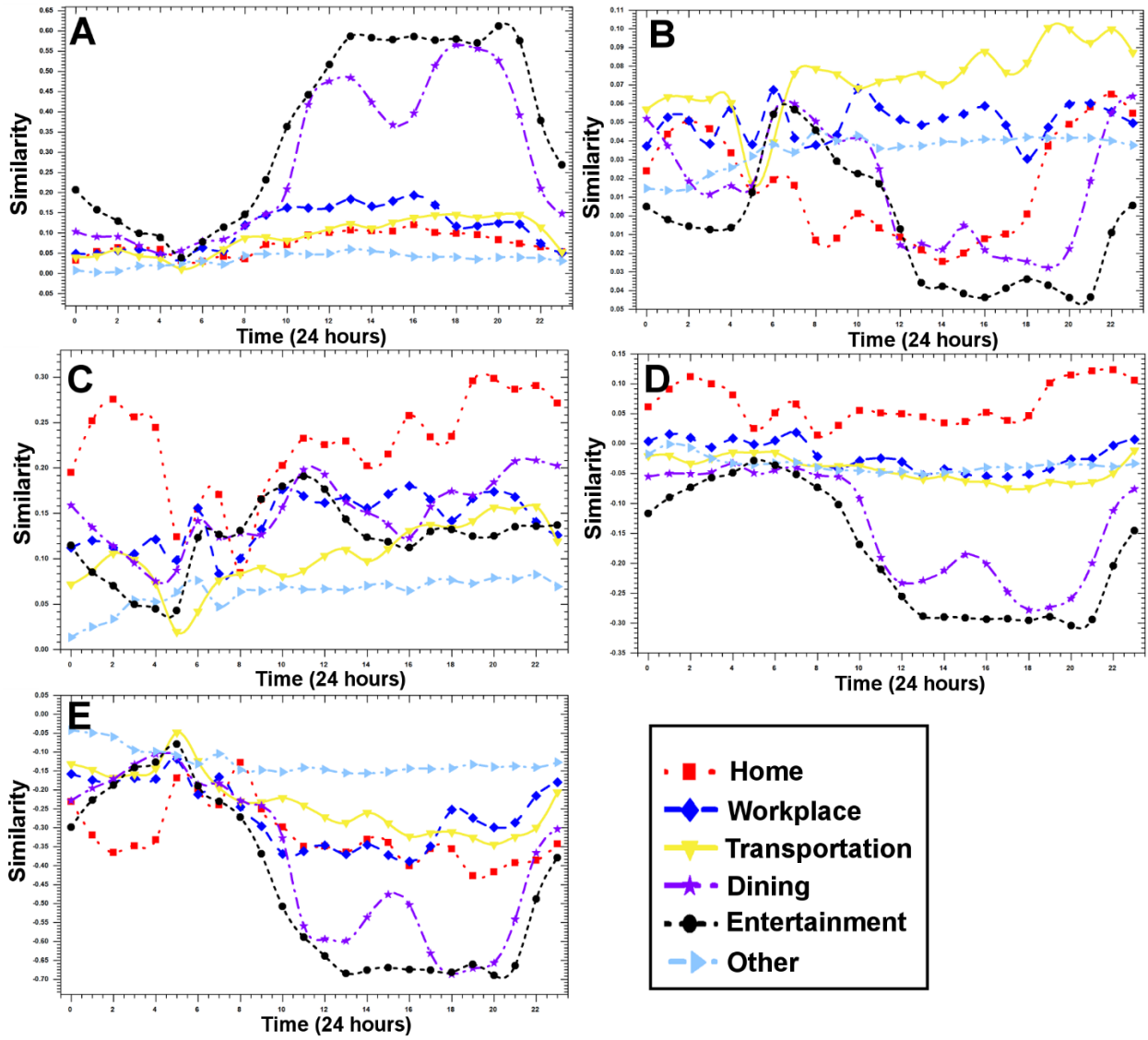

**Figure 10 The similarities among each TTDs and the centers of the five clusters.** (a) the cluster #1; (b) the cluster #2 ;(c) the cluster #3; (d) the cluster #4. (e) the cluster #5. This figure shows the similarities between each TTDs and the centers of each clusters.

In order to verify the results, we first adopt the standard deviational ellipse method to determine the directional trend of check-in dataset (the location of center is x=354344.659553, y= 3456240.891376, the ratio of the major axis to the minor axis is 1.14 and the rotation is 54.08°) comparing to the Liu et al.'s work {Liu, 2012 #11}, and it is obvious to show that the urban development of Shanghai is confined by the coastal line as well as the Yangtze River. Then, we draw nine ellipse with the with the same center and the major axis increases from 1 km to 17 km by an increment of 2 km. In each of zones, the proportions of each cluster areas are computed. Fig. 12b shows the distribution of various clusters within each zone. From inner to outer zones, the ratios of CD and APSD areas dramatically decrease while the RD and OD are significantly rise as the radius increases. Being different from the other three areas, the TD areas first increase and then decline after the radius longer than 9km. It indicates declining activity intensity gradually toward outer zones, and corresponding change in the type of region from mostly CD and APSD areas to TD area, then toward more RD and OD areas, indicating concentric form of Shanghai.

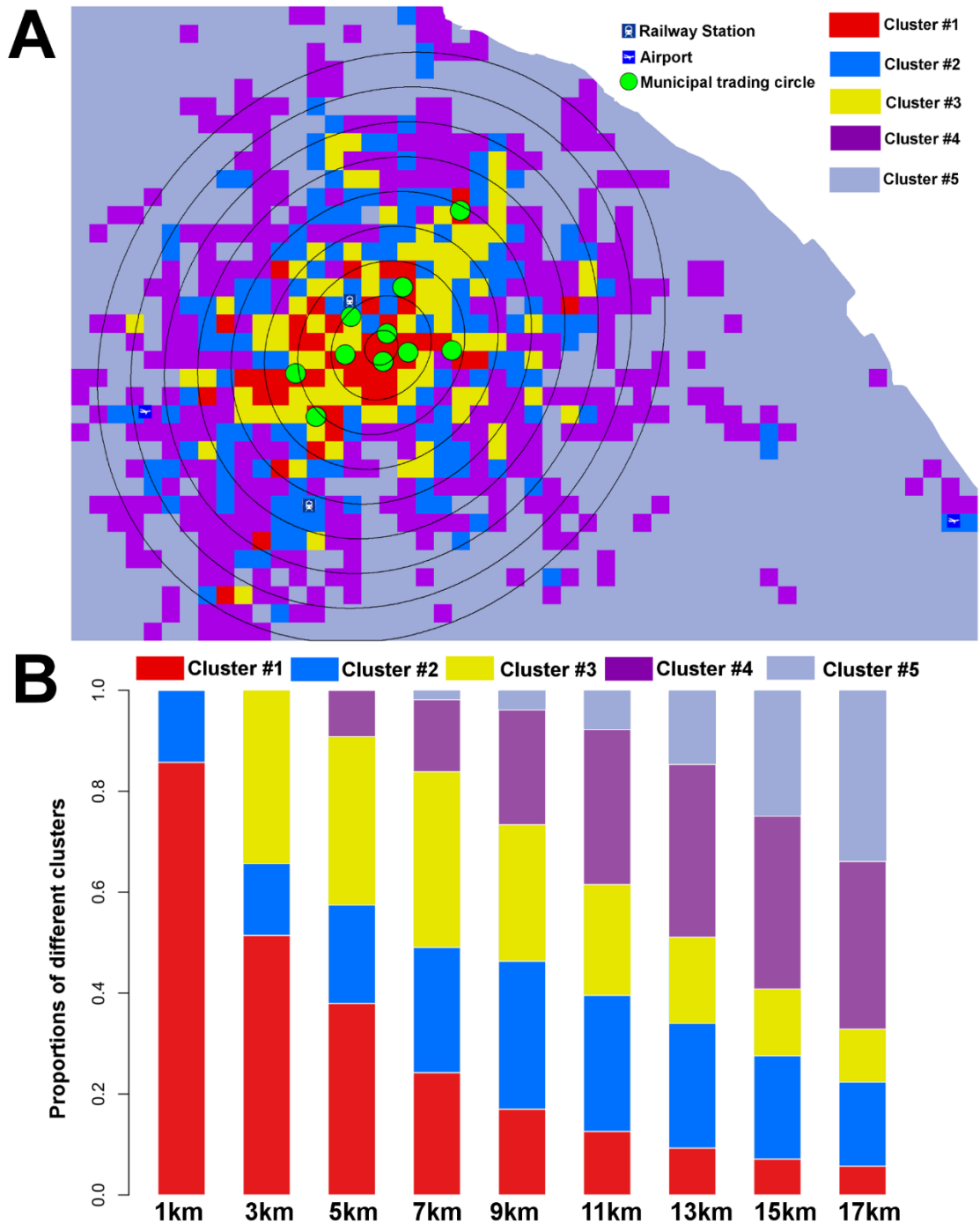

**Figure 11. The result of region aggregation and the concentric urban form revealed by the clusters**: (a) six circles with radius of 1 km, 3 km, 5km, 7km, 9 km, 11km, 13km,15km and 17 km; (b) proportions of various clusters in zones. The sum of all the clusters 'proportion equals to one in one certain radius.

The result is not only consistent with the official design, a Collection of Policies for Development in Shanghai Commercial Sector 2010, but also fits the findings of previous research well using the census and survey data [34] or the taxi data [9]. Most importantly, this result proves that how well to utilize the USTAS to infer the functional regions without any predefined knowledge and provides a clear picture of how groups of regions are associated with different travel demands at different time of day in the city.

## 5. Discussion

Most existing literature focuses on the exterior temporal rhythms of human movement to delineate different functional regions in a city, but neglects underlying interactive patterns between the functional regions and human activity due to the lack of explicit large scale activity information data. In this paper, we proposal a novel model based on a low rank approximation (LRA) to detect the functional regions using the data from about 15 million check-in records during a yearlong period in Shanghai, China. Consequently, we find a series of underlying structures, called the urban spatial-temporal activity structure (USTAS). Each one of USTAS has significant spatial and TTD characteristics. According to the top thirty ERAS, we then adopt K-means clustering algorithm to aggregate regions into five significant types based on three kinds of cluster validity indices. After analyzing the distance of each kind of region from the center of the study area, the concentric form of Shanghai can be clearly found. As a consequence, we can safely conclude that USTAS can be effectively represent the underlying structures of the Region-TTD matrix and inferring the functional regions well. Most importantly, these findings open new directions for understanding the functional regions using the LRA method. However, these work still need to be improved from two aspects in the future:

First, since we only focused on the functional regions in Shanghai in this research, we are still curious about the questions: Does the USTAS exist in any city? Are the saliences of USTAS the same among different cities, and if they are different, what are these differences try to tell us, and does these differences can reflect the economic and social structure? Hence, we will test our method on more cities to answer the above questions.

Second, in this paper, we focus on the frequency relationship of check-ins data between functional regions and TTD. We did not consider the transform structures from different TTDs, which will give us a new insight into the dynamic relationship between functional regions and TTD. However, this addition transforms information will bring in the extra dimension, which beyond the description ability of matrix. Thus, we will adopt the tensor decomposition method to model and deal with this problem.

# Reference

1. Antikainen J (2005) The concept of functional urban area. Findings of the Espon project 1.
2. Yuan J, Zheng Y, Xie X. Discovering regions of different functions in a city using human mobility and POIs; 2012. ACM. pp. 186-194.
3. Cranshaw J, Schwartz R, Hong JI, Sadeh NM. The Livehoods Project: Utilizing Social Media to Understand the Dynamics of a City; 2012.
4. Wu S, Qiu X, Usery EL, Wang L (2009) Using geometrical, textural, and contextual information of land parcels for classification of detailed urban land use. Annals of the Association of American Geographers 99: 76-98.
5. Karlsson C, Olsson M (2006) The identification of functional regions: theory, methods, and applications. The annals of regional science 40: 1-18.
6. Reades J, Calabrese F, Ratti C (2009) Eigenplaces: analysing cities using the space-time structure of the mobile phone network. Environment and Planning B: Planning and Design 36: 824-836.
7. Toole JL, Ulm M, González MC, Bauer D. Inferring land use from mobile phone activity; 2012. ACM. pp. 1-8.
8. Qi G, Li X, Li S, Pan G, Wang Z, et al. Measuring social functions of city regions from large-scale taxi behaviors; 2011. IEEE. pp. 384-388.
9. Liu Y, Wang F, Xiao Y, Gao S (2012) Urban land uses and traffic 'source-sink areas': Evidence from GPS-enabled taxi data in Shanghai. Landscape and Urban Planning 106: 73-87.
10. Pelletier M-P, Trépanier M, Morency C (2011) Smart card data use in public transit: A literature review. Transportation Research Part C: Emerging Technologies 19: 557-568.
11. Liu L, Hou A, Biderman A, Ratti C, Chen J. Understanding individual and collective mobility patterns from smart card records: A case study in Shenzhen; 2009. IEEE. pp. 1-6.
12. Calabrese F, Reades J, Ratti C (2010) Eigenplaces: segmenting space through digital signatures. Pervasive Computing, IEEE 9: 78-84.
13. Steiner RL (1994) Residential density and travel patterns: review of the literature. Transportation Research Record.
14. Kockelman KM (1997) Travel behavior as function of accessibility, land use mixing, and land use balance: evidence from San Francisco Bay Area. Transportation Research Record: Journal of the Transportation Research Board 1607: 116-125.
15. Jiang S, Ferreira Jr J, Gonzalez MC. Discovering urban spatial-temporal structure from human activity patterns; 2012. ACM. pp. 95-102.
16. Noulas A, Scellato S, Mascolo C, Pontil M (2011) An Empirical Study of Geographic User Activity Patterns in Foursquare. ICWSM 11: 70-573.
17. Cheng Z, Caverlee J, Lee K, Sui DZ (2011) Exploring Millions of Footprints in Location Sharing Services. ICWSM 2011: 81-88.
18. Wu L, Zhi Y, Sui Z, Liu Y (2014) Intra-Urban Human Mobility and Activity Transition: Evidence from Social Media Check-In Data. PloS one 9: e97010.
19. Silva TH, Melo PO, Almeida JM, Salles J, Loureiro AA. Visualizing the invisible image of cities; 2012. IEEE. pp. 382-389.
20. Sun J, Yuan J, Wang Y, Si H, Shan X (2011) Exploring space–time structure of human mobility in urban space. Physica A: Statistical Mechanics and its Applications 390: 929-942.
21. Liu Y, Sui Z, Kang C, Gao Y (2014) Uncovering Patterns of Inter-Urban Trip and Spatial Interaction from Social Media Check-In Data. PloS one 9: e86026.
22. Sun JB, Yuan J, Wang Y, Si HB, Shan XM (2011) Exploring space-time structure of human mobility in urban space. Physica A 390: 929-942.

23. Jiang S, Ferreira J, González MC (2012) Clustering daily patterns of human activities in the city. Data Min Knowl Discovery 25: 478-510.
24. Eagle N, Pentland AS (2009) Eigenbehaviors: Identifying structure in routine. Behavioral Ecology and Sociobiology 63: 1057-1066.
25. Chan KW (2007) Misconceptions and complexities in the study of China's cities: Definitions, statistics, and implications. Eurasian Geography and Economics 48: 383-412.
26. Deerwester S, Dumals ST, Furnas GW, Landauer TK, Harshman R (1990) Indexing by Latent Semantic Analysis. Journal of the American Society for Information Science 41: 391-407.
27. Ma L, Wang C, Xiao B, Zhou W. Sparse Representation for Face Recognition based on Discriminative Low-Rank Dictionary Learning; 2012.
28. Peng H, Li B, Ji R, Hu W. Salient Object Detection via Low-Rank and Structured Sparse Matrix Decomposition; 2013.
29. Candès EJ, Li X, Ma Y, Wright J (2011) Robust principal component analysis? Journal of the ACM 58: 11:11-37.
30. Jain AK (2010) Data clustering: 50 years beyond K-means. Pattern Recognition Letters 31: 651-666.
31. Aharon M, Elad M, Bruckstein A (2006) K-SVD: An Algorithm for Designing Overcomplete Dictionaries for Sparse Representation. IEEE Transactions on Signal Processing 54: 4311-4322.
32. Brun M, Sima C, Hua J, Lowey J, Carroll B, et al. (2007) Model-based evaluation of clustering validation measures. Pattern Recognition 40: 807-824
33. Halkidi M, Batistakis Y, Vazirgiannis M (2001) Journal of Intelligent Information Systems,. Journal of Intelligent Information Systems 17: 107-145.
34. Li Z, Wu F, Gao X (2007) Polarization of the global city and sociospatial differentiation in Shanghai. Scientia Geographica Sinica 27: 304.